\documentclass[12pt]{iopart}
\usepackage{amssymb}
\usepackage{bm}
\usepackage{graphicx}
\usepackage{color}
\usepackage[english]{babel}

\newcommand{\eqref}{\ref}
\newcommand{\er}{\eqref}

\begin{document}

\title[Chiral Symmetry Restoration with a Chiral Chemical Potential]
{Critical Temperature of Chiral Symmetry Restoration for Quark Matter with a Chiral Chemical Potential}

\author{M. Ruggieri}
\address{College of Physics, University of Chinese Academy of Sciences, 
Yuquanlu 19A, Beijing 100049, China.}
\ead{marco.ruggieri@ucas.ac.cn}

\author{G.~X.~Peng}
\address{College of Physics, University of Chinese Academy of Sciences, 
Yuquanlu 19A, Beijing 100049, China.}
\address{Theoretical Physics Center for Science Facilities, Institute of High Energy Physics, Beijing 100049, China.}
\ead{gxpeng@ucas.ac.cn}


\begin{abstract}
In this article we study restoration of chiral symmetry at finite temperature for quark matter
with a chiral chemical potential, $\mu_5$, by means of a nonlocal 
Nambu-Jona-Lasinio model. This model allows to introduce in the simplest way possible
a Euclidean momentum, $p_E$, dependent quark mass function which 
decays (neglecting logarithms) as $1/p_E^2$ for large $p_E$, in agreement with asymptotic behaviour
expected in QCD in presence of a nonperturbative quark condensate.
We focus on the critical temperature for chiral symmetry restoration in the chiral limit, $T_c$, versus $\mu_5$,
as well as on the order of the phase transition. We find that $T_c$ increases with $\mu_5$,
and that the transition remains of the second order for the whole range of $\mu_5$ considered.

\end{abstract}


\vspace{2pc}



\section{Introduction}
Systems with chirality imbalance, namely with a finite chiral density
$n_5=n_R-n_L$ generated by quantum anomalies, have attracted some interest in recent years.
In fact gauge field configurations with a finite winding number, $Q_W$,
can change fermions chirality according to the Adler-Bell-Jackiw anomaly~\cite{Adler:1969gk,Bell:1969ts}. 
In the context of Quantum Chromodynamics (QCD)  such nontrivial gauge field configurations with
$Q_W\neq0$ are instantons and sphalerons, the latter being produced 
copiously at high temperature~\cite{Moore:2000ara,Moore:2010jd}.
The large number of sphaleron transitions in high temperature
phase of QCD suggested the possibility to measure the Chiral Magnetic Effect
(CME)~\cite{Kharzeev:2007jp,Fukushima:2008xe} in heavy ion collisions.
The interest for mediums with a net chirality has then
spread from QCD to hydrodynamics and condensed matter systems
\cite{Son:2009tf,Banerjee:2008th,Landsteiner:2011cp,Son:2004tq,
Metlitski:2005pr,Kharzeev:2010gd,Chernodub:2015gxa,Chernodub:2015wxa,Chernodub:2013kya,
Braguta:2013loa,Li:2014bha,
Sadofyev:2010pr,Sadofyev:2010is,Khaidukov:2013sja,Kirilin:2013fqa,Avdoshkin:2014gpa}.

In order to describe systems with finite chirality in thermodynamical equilibrium,
it is customary to introduce the chiral chemical potential, $\mu_5$,
which is conjugated to $n_5$,
see~\cite{Ruggieri:2016cbq,
Gatto:2011wc,Fukushima:2010fe,Chernodub:2011fr,Ruggieri:2011xc,Yu:2015hym,Yu:2014xoa,
Braguta:2015owi,Braguta:2015zta,Braguta:2016aov,Hanada:2011jb,Xu:2015vna,
Wang:2015tia,Ebert:2016ygm,Afonin:2015cla,Andrianov:2012dj,Andrianov:2013qta} and references therein. 
Naming $\tau$ the typical time scale in which
chirality changing processes take place, it can be assumed
that $\mu_5\neq0$ describes a system in thermodynamical equilibrium with a fixed value 
of $n_5$ on a time scale much larger than $\tau$. For example in the quark-gluon
plasma phase of QCD chirality changing processes have been studied in~\cite{Manuel:2015zpa}
where it has been found that $\tau\simeq 50\div140$ fm/c in the temperature range
$T\simeq225\div500$ MeV.

An interesting problem in the context of QCD is the study of chiral symmetry restoration 
at finite temperature and $\mu_5\neq0$. 
Some previous calculations based on chiral models predicted $T_c$, the critical temperature
for chiral symmetry restoration, to decrease with $\mu_5$~\cite{Gatto:2011wc,Fukushima:2010fe,
Chernodub:2011fr,Ruggieri:2011xc,Yu:2015hym}.
On the other hand, lattice simulations have shown that $T_c$ increases 
with $\mu_5$~ \cite{Braguta:2015owi,Braguta:2015zta},
in agreement with the results obtained by solving Schwinger-Dyson equations~\cite{Xu:2015vna,Wang:2015tia}.

In this article we study chiral symmetry restoration at finite temperature
with $\mu_5\neq0$, within a Nambu-Jona-Lasinio (NJL) model~\cite{Nambu:1961tp,Nambu:1961fr,
Klevansky:1992qe,Hatsuda:1994pi,Buballa:2003qv}
with a nonlocal interaction kernel~\cite{Schmidt:1994di,Bowler:1994ir,Plant:1997jr,Blaschke:2000gd,
GomezDumm:2005hy,Frasca:2011bd,Frasca:2013kka}.
The main result of our study is that $T_c$
increases with $\mu_5$ for all the nonlocal kernels used in actual calculations.
Moreover, we discuss the order of the chiral phase transition at finite $\mu_5$: we find that
although the chiral chemical potential makes the phase transition sharper, it remains of the second 
order in the range of $\mu_5$ studied.
Both $T_c(\mu_5)$ and the absence of a critical endpoint are in agreement with the most recent
lattice results mentioned above. 

The plan of the Article is as follows. In Section 2 we briefly describe the nonlocal NJL model
we use in our calculation, presenting the several choices we do for the running dependent
mass. In Section 3 we compute the critical temperature for chiral symmetry restoration as a function
of $\mu_5$, as well as determine the order of the phase transition. 
In Section 4 we perform a small $\mu_5$ computation of the second Ginzburg-Landau (GL) 
coefficient in the free energy.
Finally in Section 5 we draw our conclusions.

\section{NJL model with momentum dependent quark mass function\label{Sec:2}}
In this Section we describe the model we use to compute the critical line for chiral symmetry restoration
in the $T-\mu_5$ plane. We use a Nambu-Jona-Lasinio (NJL) model~\cite{Nambu:1961tp,Nambu:1961fr}
(see~\cite{Klevansky:1992qe,Hatsuda:1994pi,Buballa:2003qv} for reviews)
with a nonlocal interaction kernel inspired by the Instanton Liquid picture 
of the QCD vacuum~\cite{Plant:1997jr,Schmidt:1994di,Bowler:1994ir,Blaschke:2000gd,GomezDumm:2005hy},
see \cite{Schafer:1996wv} for a review,
which has the advantage to introduce in the simplest way possible a Euclidean momentum
dependent quark mass function that agrees with QCD~\cite{Politzer:1976tv,Gasser:1982ap}.

\subsection{Thermodynamic potential}
In the nonlocal NJL model we use in this study the lagrangian density is given by
\begin{equation}
{\cal L} =  \bar\psi\left(
i\partial^\mu\gamma_\mu  + \mu_5\bar\psi\gamma_0\gamma_5\psi
\right)\psi + {\cal L}_4,
\label{eq:Snl1}
\end{equation}
with $\psi$ being a quark field with Dirac, color and flavor indices.  In this equation
$\mu_5$ is the chiral  chemical potential, and its conjugated
quantity is the chiral charge density, $n_5\equiv n_R - n_L$: a finite $\mu_5$ induces
a chiral density in the system, and in general the relation between $n_5$ and $\mu_5$
has to be computed numerically within some model, see for 
example~\cite{Fukushima:2010fe,Chernodub:2011fr}.

In Eq.(\er{eq:Snl1}) ${\cal L}_4$ corresponds to the interaction term, namely
\begin{equation}
{\cal L}_4 = G 
\int d^4y d^4z F^\star(y-x)F(z-x) \bar\psi(y)\psi(z).
\label{eq:Snl2}
\end{equation}
The interaction term in Eq. (\er{eq:Snl2})
is formally equivalent to a local NJL interaction,
\begin{equation}
{\cal L}_4 = G (\bar\Psi(x)\Psi(x))^2,
\end{equation}
written in terms of the dressed quark fields
\begin{equation}
\Psi(x) \equiv \int d^4y F(y-x) \psi(y).
\label{eq:dressing}
\end{equation}

Chiral symmetry is spontaneously broken by the  interaction in Eq.~(\ref{eq:Snl2});
this leads to a nonvanishing expectation value of the
dressed quark field operator
\begin{equation}
\sigma \equiv G\langle\bar\Psi(x)\Psi(x)\rangle\neq0.
\label{eq:condensate}
\end{equation}
Working at finite temperature $T$ within the well established imaginary time formalism,
the thermodynamic potential per unit volume can be written as
\begin{eqnarray}
\Omega &=& \frac{\sigma^2}{G}
-N_c N_f T\sum_n\int\frac{d^3\bm p}{(2\pi)^3} \log \beta^4 
(\omega_n^2 + E_+^2)(\omega_n^2 + E_-^2),
\label{eq:TP2}
\end{eqnarray}
where $\beta=1/T$ and we have defined
\begin{equation}
E_\pm^2 = (p \pm \mu_5)^2 + M(\omega_n,\bm p)^2.
\label{eq:ip1}
\end{equation}
Here $\omega_n = \pi T(2n+1)$ corresponds to the fermionic Matsubara frequency and $M(\omega_n,\bm p) $
denotes the quark mass function to be specified later.

\subsection{Quark mass functions}
In Eq.~(\ref{eq:ip1}) we have introduced the quark mass function
\begin{equation}
M(p) \equiv -2\sigma {\cal C}(p),
\label{eq:mm2}
\end{equation}
with
${\cal C}(p) \equiv F^2(p)$ and $F(p)$ corresponding to the Fourier transform of the form factor $F$ in 
Eq. (\er{eq:dressing}). 
The above equation agrees with the results from one-gluon exchange inspired 
models~\cite{Hell:2008cc,Hell:2009by,Frasca:2011bd,Frasca:2013kka,Langfeld:1996rn,Langfeld:1996ac}.
Nonlocal models mimic 
the constituent quark mass function of QCD in presence of 
spontaneous chiral symmetry breaking~\cite{Politzer:1976tv,Langfeld:1996rn} for large $p_E$.
In this work we assume several specific functional forms for $M(p)$ in
Eq. (\er{eq:mm2}).

A class of form factors that we use have the form
\begin{eqnarray}
{\cal C}(p_E^2) &=& \theta(\Lambda^2 - p_E^2)  + \theta(p_E^2 - \Lambda^2)
\frac{\Lambda^2 (\log\Lambda^2/\Lambda_{QCD}^2)^{\gamma}}
{p_E^2 (\log p_E^2/\Lambda_{QCD}^2)^\gamma}.
\label{eq:LL2}
\end{eqnarray}
For the exponent $\gamma$ in Eq.(\er{eq:LL2}) we consider here three cases:
$\gamma=0$ for its simplicity; $\gamma=1$ following~\cite{Hell:2008cc,Hell:2009by};
finally $\gamma=1-d_m$, inspired by the quark mass function 
derived by Politzer~\cite{Politzer:1976tv}, where $d_m=12/29$ corresponds to the anomalous
mass dimension for $N_f=2$.

We also consider form factors that connect smoothly the infrared and the ultraviolet $p_E$ domains.
In particular, we consider a Yukawa-type form factor, namely~\cite{Frasca:2011bd,Frasca:2013kka}
\begin{equation}
{\cal C}(p_E^2) = 
\frac{\Lambda^2}
{p_E^2 + \Lambda^2};
\label{eq:LL2y}
\end{equation} 
then we consider a form factor inspired by the Instanton Liquid Model (ILM)
of the QCD vacuum, namely~\cite{Schafer:1996wv}
\begin{eqnarray}
{\cal C}(p_E^2) =\frac{ d^2 p_E^2}{4}
\left\{
\frac{d}{dx}[I_0(x)K_0(x)-I_1(x)K_1(x)]
\right\}^2,
\label{eq:LL2ilm}
\end{eqnarray}  
where $d$ corresponds to the typical instanton size $d\approx0.36$ fm and $x=|p_E|d/2$.
Finally we consider a nonlocal kernel used in nonlocal NJL model studies~\cite{Hell:2008cc,Hell:2009by}, namely
\begin{eqnarray}
{\cal C}(p_E^2) =\theta(\Lambda^2-p_E^2)e^{-p_E^2 d^2/2} +\theta(p_E^2-\Lambda^2)
\frac{\Lambda^2 (\log\Lambda^2/\Lambda_{QCD}^2)}
{p_E^2 (\log p_E^2/\Lambda_{QCD}^2)}e^{-\Lambda^2 d^2/2},
\label{eq:LL2c}
\end{eqnarray}
where $d$ corresponds to the instanton size used in Eq. (\er{eq:LL2ilm}) and $\Lambda=O(1)$ GeV.
Equation (\er{eq:LL2c}) offers a smooth version of the form factor in Eq. (\er{eq:LL2}) with $\gamma=1$.
 
\section{The critical temperature and the order of the phase transition}
In this Section we compute the critical line for chiral symmetry restoration as a function of the
chiral chemical potential, both within a GL expansion of the thermodynamic potential
in Eq.(\er{eq:TP2}) and within numerical calculations using the full potential. 

\subsection{Ginzburg-Landau expansion}
Close to a second order phase transition we can write Eq.(\er{eq:TP2}) as
\begin{equation}
\Omega - \Omega_0= \frac{\alpha_2}{2}\sigma^2 + \frac{\alpha_4}{4}\sigma^4 +O(\sigma^6), 
\label{eq:25}
\end{equation}
where we have subtracted the thermodynamic potential
at $\sigma=0$, namely $\Omega_0$; $\alpha_2$ and $\alpha_4$ can be computed by taking the derivatives of 
$\Omega$ with respect to $\sigma$ at $\sigma=0$. We find
\begin{eqnarray}
\alpha_2 &=&\frac{2}{G} -N_c N_f T\sum_n\int\frac{d^3\bm p}{(2\pi)^3} 
{\cal C}^2(\omega_n,\bm p)\nonumber\\
&&\times\frac{16(\bm p^2 + \omega_n^2 +\mu_5^2)}
{[\omega_n^2 + (p-\mu_5)^2][\omega_n^2 + (p+\mu_5)^2]},
\label{eq:al2}
\end{eqnarray}
and
\begin{eqnarray}
\alpha_4 &=&-N_c N_f T\sum_n\int\frac{d^3\bm p}{(2\pi)^3} 
{\cal C}^4(\omega_n,\bm p)\nonumber\\
&&\frac{-384[\bm p^4 +2\bm p^2(\omega_n^2 +3\mu_5^2) +(\mu_5^2 + \omega_n^2)^2]}
{[\omega_n^2 + (p-\mu_5)^2]^2[\omega_n^2 + (p+\mu_5)^2]^2}.
\label{eq:al4}
\end{eqnarray}
The nontrivial solution of the gap equation, $\partial\Omega/\partial\sigma=0$, for $T\le T_c$ is given by
\begin{equation}
\sigma^2(T,\mu_5) = -\frac{\alpha_2(T,\mu_5)}{\alpha_4(T,\mu_5)},
\label{eq:gsl}
\end{equation}
and the critical temperature is defined by the condition $\alpha_2(T,\mu_5)=0$.

\begin{figure}[t!]
\begin{center}
\includegraphics[width=7cm]{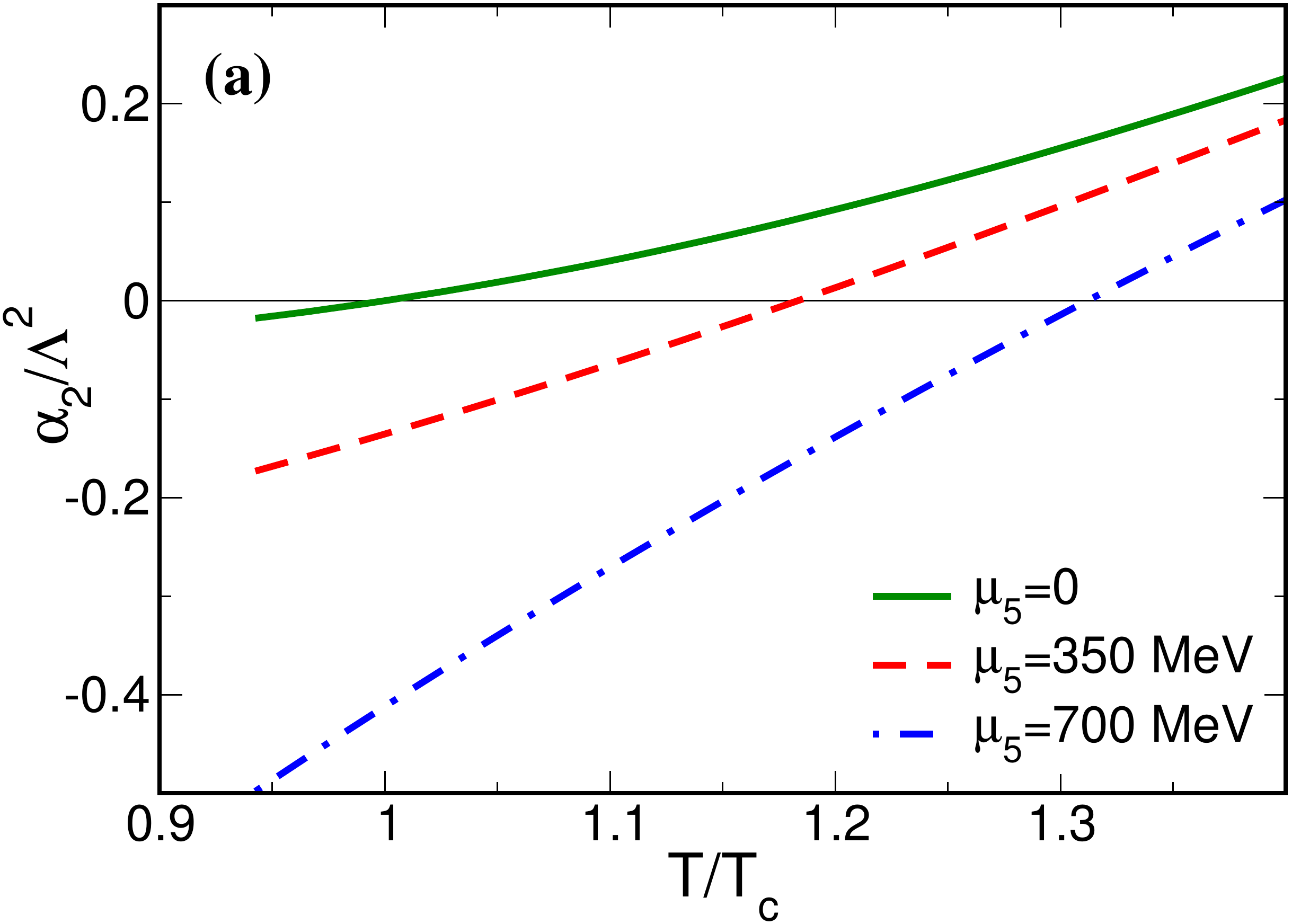}\\
\includegraphics[width=7cm]{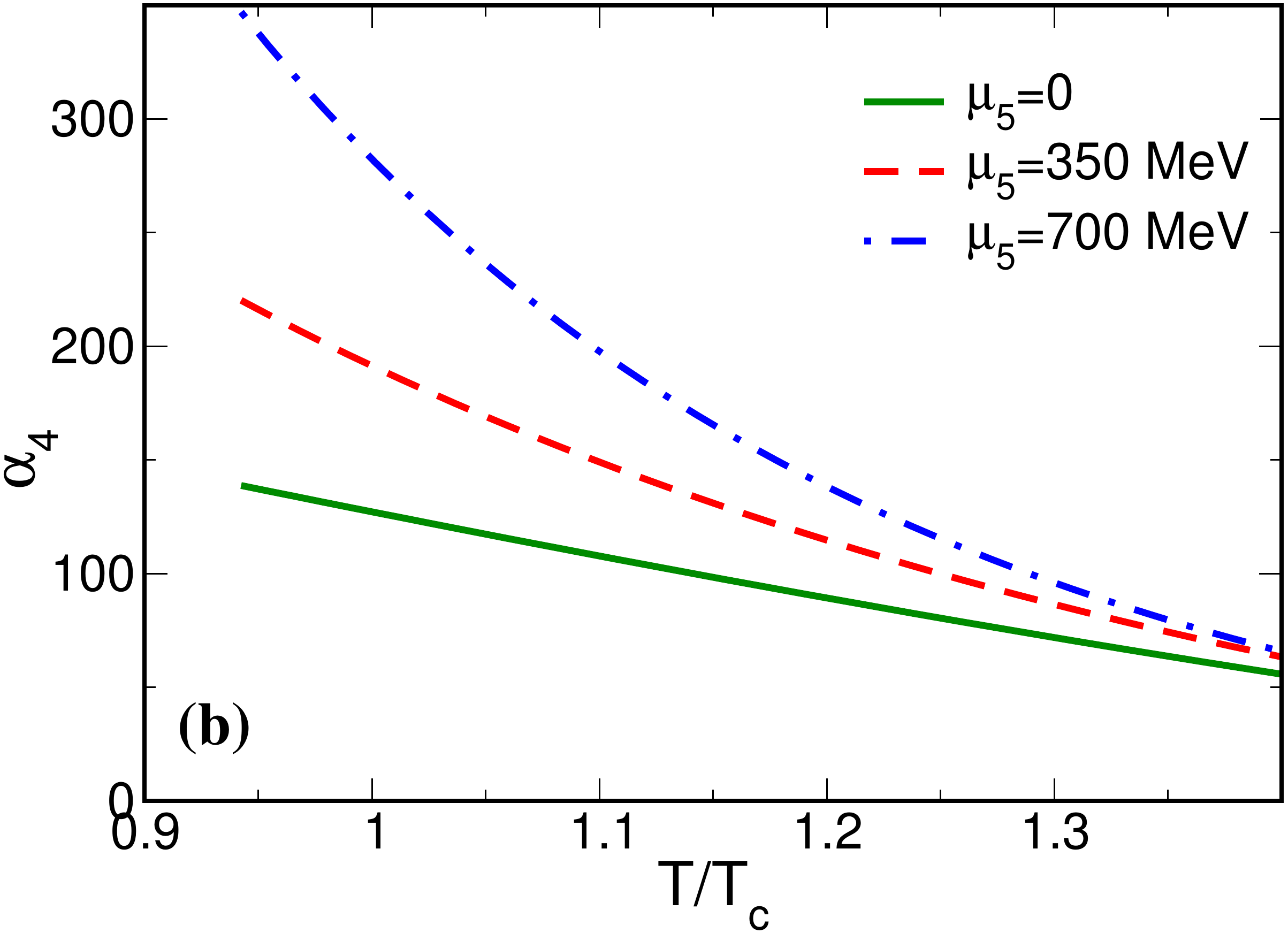}\\
\includegraphics[width=7cm]{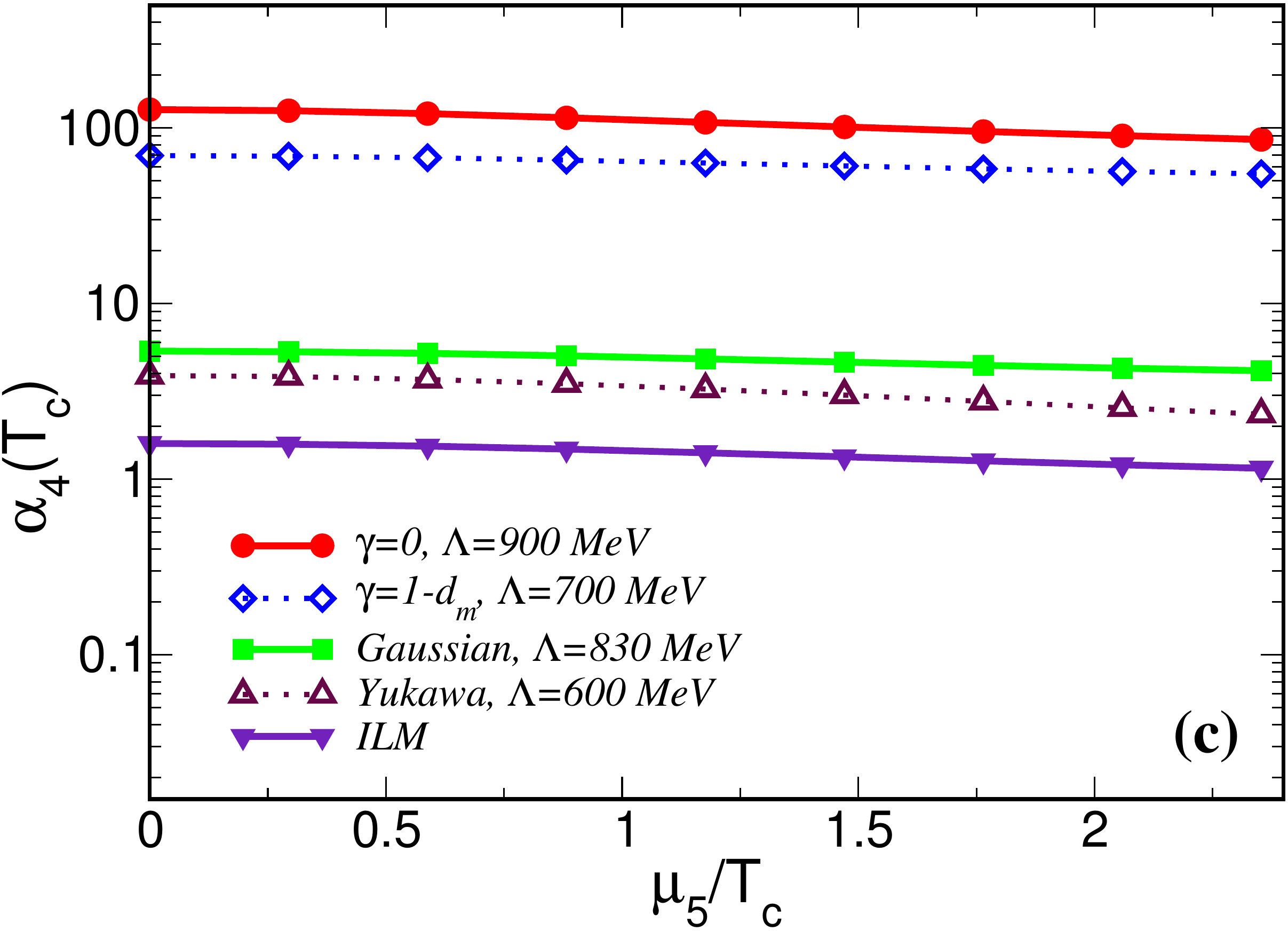}
\end{center}
\caption{\label{Fig:5}{\it (a).} Coefficient $\alpha_2$ in units of the parameter $\Lambda^2$
as a function of temperature, for three different values of $\mu_5$. 
For each value of $\mu_5$ the critical temperature $T_c$ corresponds to $\alpha_2(T_c)=0$.
{\it (b).} Coefficient $\alpha_4$ 
versus temperature, for three different values of $\mu_5$. 
(a) and (b) refer to
the nonlocal model with mass function given by Eqs. (\er{eq:LL2}) and (\er{eq:mm2}) with $\gamma=0$
and $\Lambda=900$ MeV.
{\it (c).} Evolution of the coefficient $\alpha_4$ computed at $T=T_c(\mu_5)$ versus $\mu_5$
for several models.}
\end{figure}
In Fig.~\ref{Fig:5}a we plot the coefficient $\alpha_2$ in units of the parameter $\Lambda^2$
as a function of temperature, for three different values of $\mu_5$. 
For each value of $\mu_5$ the critical temperature $T_c$ corresponds to $\alpha_2(T_c)=0$.
We show data for the nonlocal model with mass function given by Eqs. (\er{eq:LL2}) 
and (\er{eq:mm2}) with $\gamma=0$ and $\Lambda=900$ MeV; for other models we obtain qualitatively
the same results. We notice that increasing $\mu_5$ results in an increasing critical temperature. 
We also notice that the slope of $\alpha_2$ at $T=T_c$ increases with $\mu_5$. Together
with the behaviour of $\alpha_4$ discussed below, this is a signature of the phase transition
becoming sharper with $\mu_5$.

In Fig.~\ref{Fig:5}b we plot the coefficient $\alpha_4$ 
versus temperature, for the same values of $\mu_5$ shown in Fig.~\ref{Fig:5}a of the figure. 
We notice that for any value of $\mu_5$ the coefficient $\alpha_4$ decreases in magnitude,
but it is always positive at the critical temperature meaning the phase transition is a second order one.
We also notice that the magnitude of $\alpha_4$ at $T=T_c(\mu_5)$ decreases compared to the
case $\mu_5=0$, implying that the phase transition becomes sharper with increasing $\mu_5$.
In fact because of Eq.~(\ref{eq:gsl}) we can write the solution of the gap equation for $T\approx T_c$ as
\begin{equation}
\sigma^2 =  -\frac{1}{\alpha_4(T_c)}
\left.\frac{d\alpha_2}{dT}\right|_{T=T_c}(T-T_c) + O[(T-T_c)^2],
\end{equation} 
then the slope of the condensate at the critical temperature is given by
\begin{equation}
\left.\frac{d\sigma^2}{dT}\right|_{T=T_c} =  -\frac{1}{\alpha_4(T_c)}
\left.\frac{d\alpha_2}{dT}\right|_{T=T_c},
\end{equation}
which becomes larger as we increase $\mu_5$ because $\alpha_4$ decreases and the slope of $\alpha_2$
at the critical temperature increases with $\mu_5$. Our conclusion is that within the range of $\mu_5$ explored 
in our study, we have a firm signal that the phase transition becomes sharper as $\mu_5$ is increased,
but there is no critical endpoint because $\alpha_4$ does not change sign at the critical temperature.
This is in agreement with lattice simulations~\cite{Braguta:2015owi,Braguta:2015zta}
but it is in disagreement with previous model studies which used a different regularization 
scheme~\cite{Gatto:2011wc,Chernodub:2011fr,Fukushima:2010fe,Ruggieri:2011xc},
showing how the existence of the critical point in the phase diagram is very sensitive on the
regularization prescription, in fact already noticed in~\cite{Yu:2015hym}.
Finally in Fig.~\ref{Fig:5}c we plot the coefficient $\alpha_4$ at $T=T_c(\mu_5)$ for
several models. We notice that although the numerical value of $\alpha_4$ strongly depends on the model,
we find that it is always positive at $T=T_c$.

\subsection{Beyond Ginzburg-Landau expansion}

In the previous subsection
we have discussed results obtained within a GL expansion of the thermodynamic potential,
see Eq.~(\ref{eq:25}). The GL expansion is a useful tool because it allows to study the analytical behaviour
of the thermodynamic potential close to the phase transition.
As long as we are interested to the critical temperature at a second order phase transition
the GL expansion is equivalent to use the full thermodynamic potential in Eq.~(\ref{eq:TP2}):
as a matter of fact, it requires a straightforward calculation to verify that
the gap equation $\partial\Omega/\partial\sigma=0$ at $T=T_c$ obtained from
Eq.~(\ref{eq:25}) coincides with the GL gap equation at $T=T_c$, that is $\alpha_2=0$ 
where $\alpha_2$ is the second order
GL coefficient in Eq.~(\ref{eq:25}). 
On the other hand, for a first order phase transition the GL expansion is not reliable because
the value of the condensate at $T=T_c$ might not be small compared to $T$, and
the use of the expansion in Eq.~(\ref{eq:25}) might be doubtful.

A natural question therefore arises, namely if the above results, in particular the order of the 
phase transition, are a mere consequence of the
GL expansion or if they are in agreement with those that would be obtained using the
gap equation derived from the full thermodynamic potential. 
In the previous Section  
we have first computed the temperature at which
$\alpha_2=0$, identifying this with $T_c$, then we have computed $\alpha_4$ at $T=T_c$
checking its sign: we have then concluded that being $\alpha_4>0$ the phase transition
is always of the second order regardless the value of $\mu_5$; within the GL
approximation $\alpha_4<0$ would have been a signal of a first order phase transition. 
The purpose of this Section is to 
check the results of the previous Section 
going beyond the GL expansion of Eq.~(\ref{eq:25}). To this end we compute the
condensate defined in Eq.~(\ref{eq:condensate}) by solving the gap equation $\partial\Omega/\partial\sigma=0$
with $\Omega$ defined in Eq.~(\ref{eq:TP2}).

\begin{figure}[t!]
\begin{center}
\includegraphics[width=7cm]{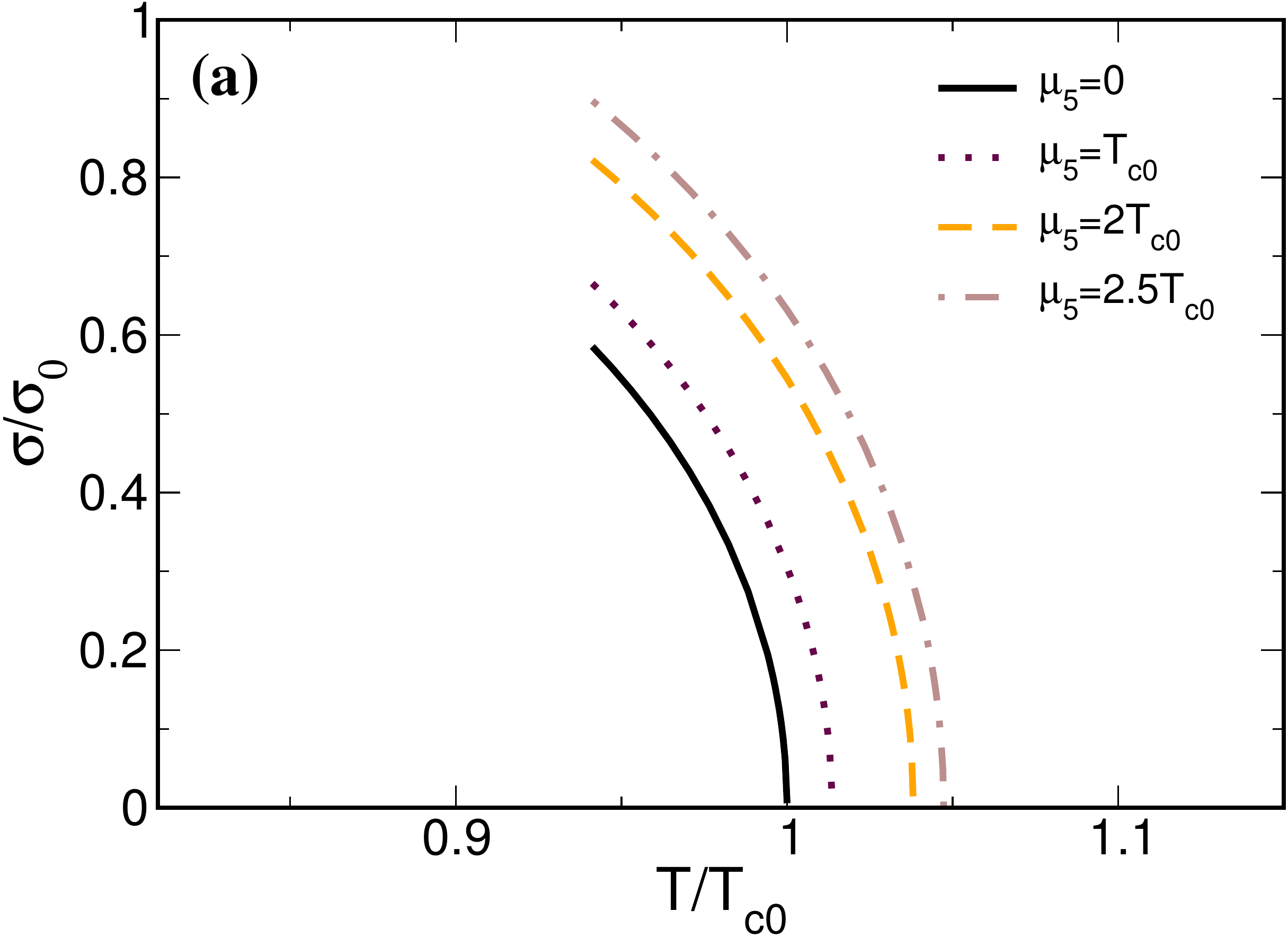}~~~~~
\includegraphics[width=7cm]{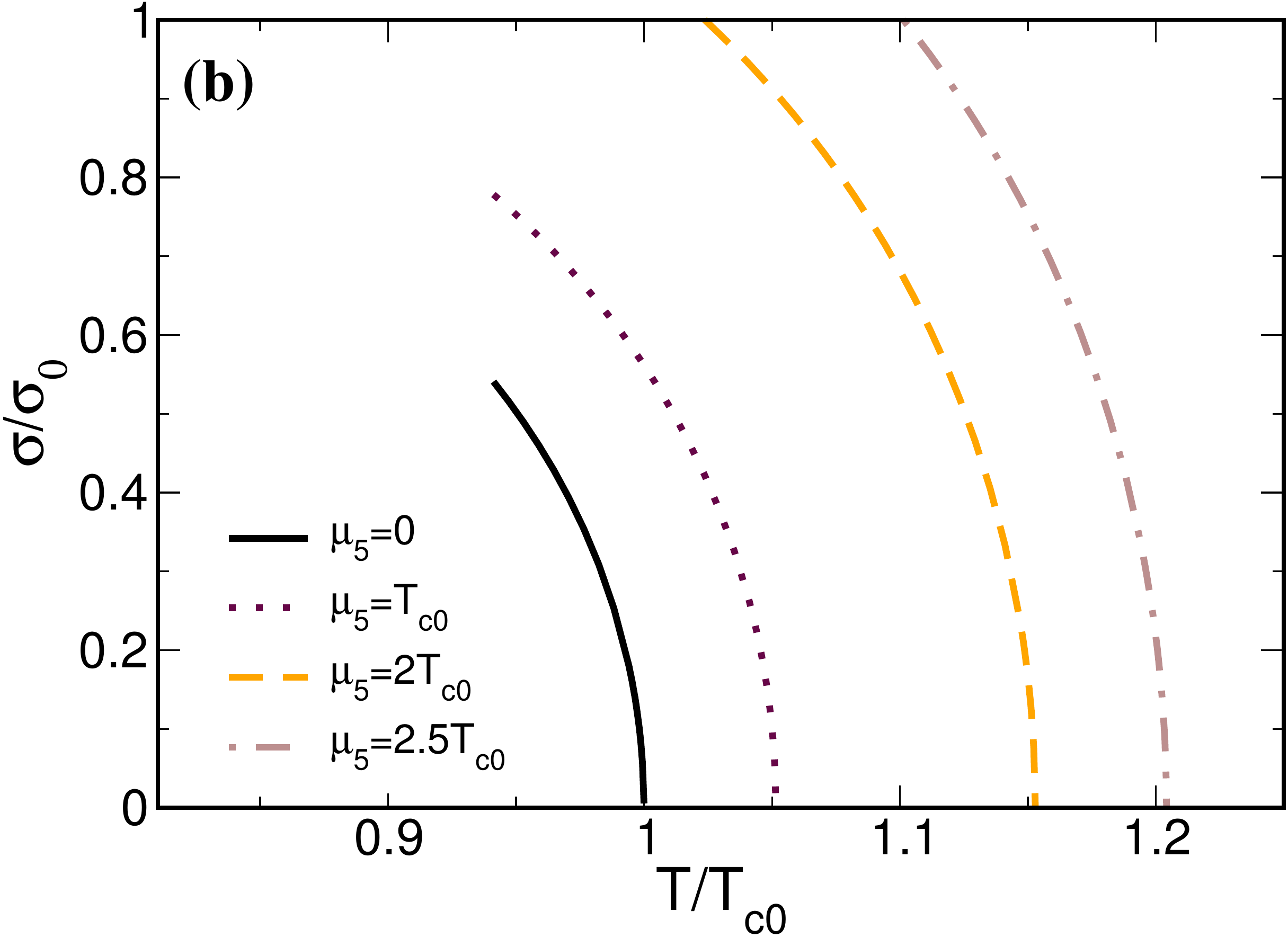}
\end{center}
\caption{\label{Fig:6b}{\it (a).} NJL condensate, defined in Eq.~(\ref{eq:condensate}),
versus temperature for several values of $\mu_5$, for the case of the Gaussian
form factor of Fig.~\ref{Fig:6}. 
{\it (b).} Condensate for the case of the Yukawa form factor with $\Lambda=900$ MeV in Fig.~\ref{Fig:6}. 
In both panels $T_{c0}$ corresponds to the critical temperature
at $\mu_5=0$, while $\sigma_0$ denotes the condensate
at $T=0$ and $\mu_5=0$. For both form factors we plot data for several values of $\mu_5$:
in particular black solid lines correspond to $\mu_5=0$, maroon dotted lines to $\mu_5=T_{c0}$,
orange dashed lines to $\mu_5=2T_{c0}$, finally brown dot-dashed lines to $\mu_5=2.5T_{c0}$.}
\end{figure}

In Fig.~\ref{Fig:6b} we plot the condensate versus temperature 
for two of the nonlocal NJL models mentioned in Fig.~\ref{Fig:6}, namely the Gaussian model
(a) and the Yukawa model with $\Lambda=900$MeV (b). We have checked that for the other models
we obtain similar results. In the figure $T_{c0}$ corresponds to the critical temperature
at $\mu_5=0$, while $\sigma_0$ denotes the condensate
at $T=0$ and $\mu_5=0$. For both form factors we plot data for several values of $\mu_5$:
in particular black solid lines correspond to $\mu_5=0$, maroon dotted lines to $\mu_5=T_{c0}$,
orange dashed lines to $\mu_5=2T_{c0}$, finally brown dot-dashed lines to $\mu_5=2.5T_{c0}$.
For both cases we have zoomed to the temperature range close to the critical temperature which is
the one relevant for our study.

Data shown in Fig.~\ref{Fig:6b} confirm the results obtained within the GL expansion and presented
in the previous subsection.
In fact, the critical temperature is found to increase with $\mu_5$. 
Moreover the condensate vanishes smoothly
with increasing temperature, meaning the phase transition is of the second order
(a first order phase transition would appear as a discontinuity in the condensate, which we do not find
for all the values of $\mu_5$ explored here). 
We thus can conclude that the main results of the our
study, with particular regard to the absence of a first order 
phase transition line in the $\mu_5-T$ plane, 
are not a mere consequence of the GL expansion Eq.~(\ref{eq:25}).
The presence of a first order phase transition line, found in NJL and quark-meson model calculations
\cite{Gatto:2011wc,Fukushima:2010fe,Chernodub:2011fr,Ruggieri:2011xc}
but not found in the nonlocal NJL model calculations,
appears thus to be model dependent, in agreement with what anticipated in \cite{Yu:2015hym}.

\subsection{The critical line in the $\mu_5-T$ plane}

\begin{figure}[t!]
\begin{center}
\includegraphics[width=7cm]{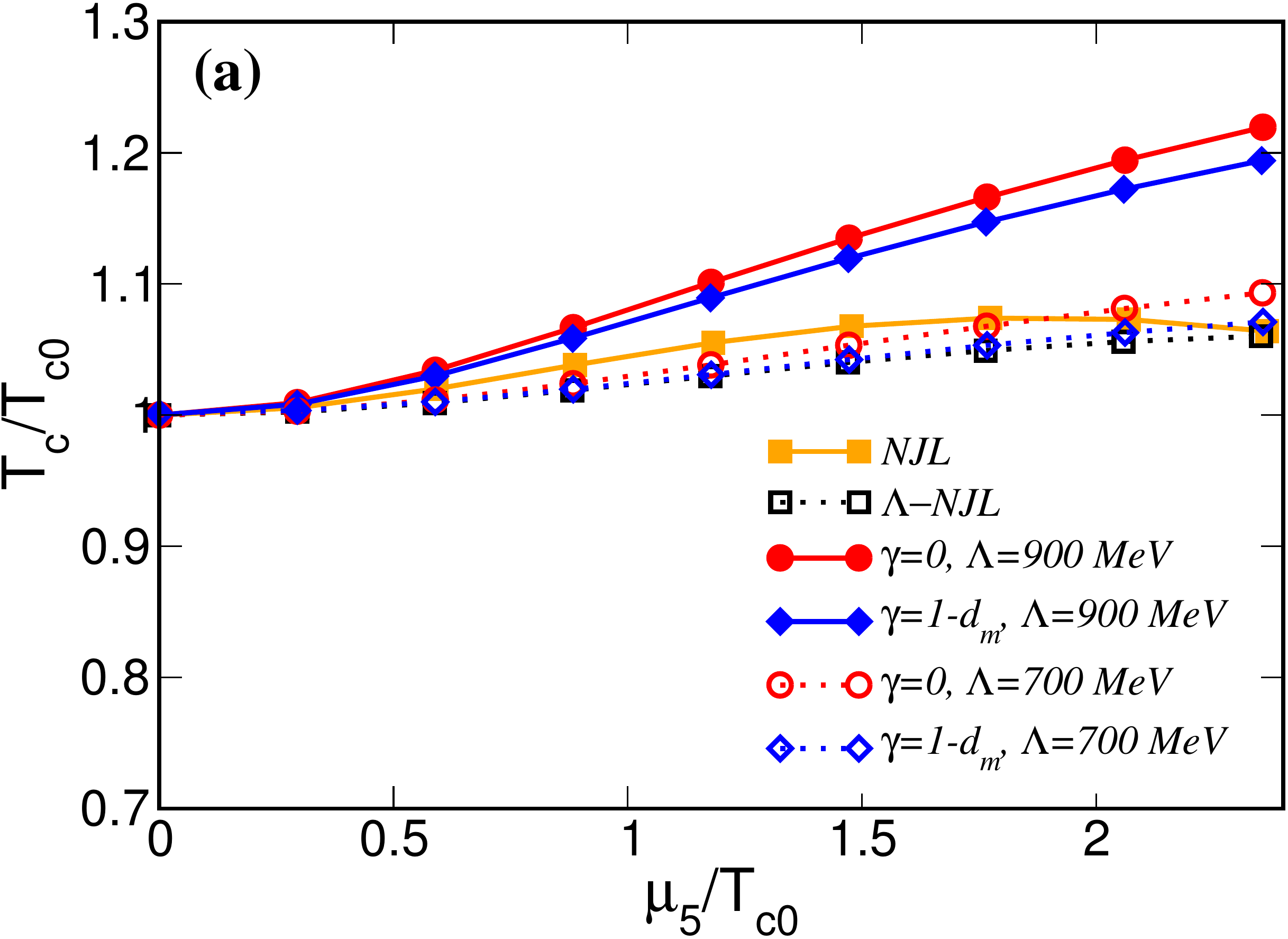}~~~~~
\includegraphics[width=7cm]{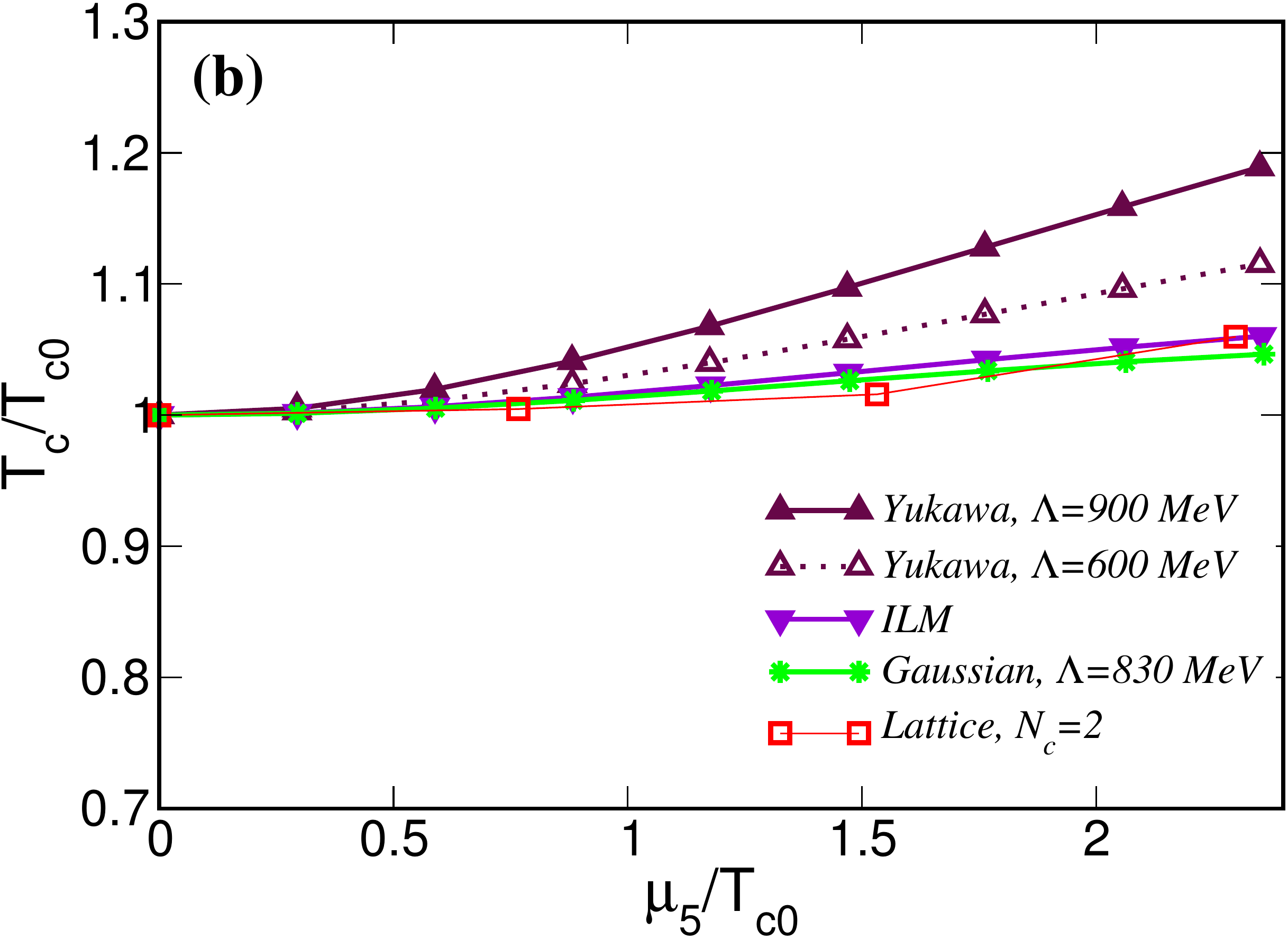}
\end{center}
\caption{\label{Fig:6}{\it (a).} Critical temperature for chiral symmetry restoration versus $\mu_5$
for several running mass models described in the text. Squares correspond to NJL and $\Lambda-$NJL
models with a 4D sharp cutoff $\Lambda=900$ MeV.
Circles correspond to mass function given by Eqs. (\er{eq:LL2}) and (\er{eq:mm2}) with $\gamma=0$,
for two different values of $\Lambda$; diamonds correspond to the same mass function with 
$\gamma=1-d_m$. 
{\it (b).} Critical temperature for chiral symmetry restoration versus $\mu_5$
for several running mass models described in the text. Data with triangles pointing upwards 
correspond to a Yukawa-like form factor in Eq. (\er{eq:LL2y}) with two values of $\Lambda$. 
Data denoted by triangles pointing downwards correspond to the Instanton Liquid Model (ILM) form factor
in Eq. (\er{eq:LL2ilm}).
Stars correspond to the nonlocal form factor in Eq. (\er{eq:LL2c}).
Lattice data for $N_c=2$ have been adapted from~\cite{Braguta:2015zta}.
In both panels $T_{c0}$ corresponds to critical temperature at $\mu_5=0$.}
\end{figure}

In Fig.~\ref{Fig:6} we plot the critical temperature versus $\mu_5$ for the nonlocal models described in the text.
In the figure $T_{c0}$ denotes the critical temperature at $\mu_5=0$.
In Fig.~\ref{Fig:6}a we collect the results for the sharp models described in the text.
Circles correspond to mass function given by Eqs. (\er{eq:LL2}) and (\er{eq:mm2}) with $\gamma=0$,
for two different values of $\Lambda$; diamonds correspond to the same mass function with 
$\gamma=1-d_m$. 
In Fig.~\ref{Fig:6}a we have also shown the results for two 
local NJL models. In particular, we denote by squares the results for a standard local NJL model with a 4-dimensional 
sharp cutoff on the vacuum term and no cutoff on the thermal part of the free energy; moreover,
empty squares correspond to a model dubbed $\Lambda-$NJL, in which there is a 4D sharp cutoff
both on the vacuum and on the thermal contribution to the gap equation. In both cases $\Lambda=900$ MeV.
In Fig.~\ref{Fig:6}b we plot the critical temperature
for smooth form factors. In particular data with triangles pointing upwards 
correspond to a Yukawa-like form factor in Eq. (\er{eq:LL2y}) with two values of $\Lambda$. 
Data denoted by triangles pointing downwards correspond to the Instanton Liquid Model (ILM) form factor
in Eq. (\er{eq:LL2ilm}).
Finally stars correspond to the nonlocal form factor in Eq. (\er{eq:LL2c}).
In both panels both temperature and chemical potential are measured in units of the critical
temperature  at $\mu_5=0$.
In each calculation we have fixed the value of the parameter $\Lambda$ in the form factor, then we have tuned
the NJL coupling constant $G$ in order to obtain $T_{c0}=170$ MeV for any model.

The results in Fig.~\ref{Fig:6} show that for all the nonlocal models studied in this article the critical temperature
increases with $\mu_5$.
For large values of $\mu_5$ the results shown in Fig.~\ref{Fig:6}  should be not considered
very reliable because we have neglected a possible backreaction on the nonlocal interaction kernel
due to $\mu_5$. For the case of local models, we find that the $\Lambda-$NJL model still predicts
$T_c$ increases with $\mu_5$, at least up to values of $\mu_5$ of the order of $\Lambda$.
This is in agreement with the previous analysis of~\cite{Yu:2015hym} where a $\Lambda-$NJL
with a 3-dimensional cutoff has been considered. For the NJL model result in Fig.~\ref{Fig:6}
we find that $T_c$ increases with $\mu_5$ for small values of $\mu_5$, in agreement with 
a small $\mu_5$ analysis presented in the following section.

A detailed comparison with lattice data~\cite{Braguta:2015owi,Braguta:2015zta} is premature because 
those data have not been obtained in the chiral limit; moreover, some data on the lattice correspond to $N_c=2$
QCD while here we consider $N_c=3$. However, we can at least compare the magnitude
of the increase of the critical temperature obtained within the nonlocal models and within the lattice simulations.
In Fig.~\ref{Fig:6} we show lattice results for $T_c(\mu_5)$
for $N_c=2$ adapted from Ref.~\cite{Braguta:2015zta} in which the critical temperature at $\mu_5=0$
is $T_{c0}=195.8\pm0.4$ MeV.  We find that among the models considered here, the ones with Gaussian
ILM form factors, respectively Eqs. (\er{eq:LL2c}) and~(\eqref{eq:LL2ilm}), better reproduce the magnitude
of the variation of the critical temperature with $\mu_5$.

\section{Small $\mu_5$ analysis}
Since the phase transition is of the second order we can use the GL expansion, see Eq.~(\ref{eq:25}),
to investigate in more detail the relation between $\mu_5$ and $T_c$ within the model at hand.
In particular, we perform in this section a small $\mu_5$ analysis of the 
coefficient $\alpha_2$ in Eq.~(\ref{eq:25}) to
enlighten the differences between local and nonlocal NJL models at finite $\mu_5$.

\subsection{The coefficient $\alpha_{2,2}$ and $T_c$ versus $\mu_5$ for $\mu_5/T\ll1$}
We expand
\begin{equation}
\alpha_2 = \alpha_{2,0} + \mu_5^2\alpha_{2,2}.
\end{equation}
The above equation allows to compute, to the lowest order in $\mu_5/ T$,
the shift of the critical temperature due to $\mu_5$:
\begin{equation}
\delta T_c = -\frac{\alpha_{2,2}(T_c^0)}{a}\mu_5^2,
\label{eq:kl}
\end{equation}
where $T_c^0$ corresponds to the critical temperature at $\mu_5=0$ and
$a\equiv d\alpha_{2,0}/dT$ at $T=T_c^0$. The quantity $a$ depends on the specific model used
but it is positive by definition because 
$\alpha_{2,0}$ is negative for $T<T_c$ and positive for $T>T_c$, thus the sign
of $\delta T$ in Eq.~\eqref{eq:kl} is determined only by the sign of $\alpha_{2,2}$.
A straightforward computation starting from Eq.(\er{eq:TP2}) shows that
\begin{eqnarray}
\alpha_{2,2} &=& -4N_c N_f T \sum_n \int\frac{d^3\bm p}{(2\pi)^3}{\cal C}^2(\omega_n,\bm p)
\frac{2(3\bm p^2 - \omega_n^2)}{(\bm p^2 + \omega_n^2)^3},
\label{eq:asd}
\end{eqnarray}
where ${\cal C}$ is the non local interaction kernel. 
Once $\alpha_{2,2}$ is known, the critical temperature versus $\mu_5$ can be
computed as
\begin{equation}
T_c(\mu_5) =T_c^0\left[1 -\frac{\alpha_{2,2}(T_c^0)}{a T_c^0}\mu_5^2\right].
\label{eq:kl2}
\end{equation}

\begin{figure}[t!]
\begin{center}
\includegraphics[width=7cm]{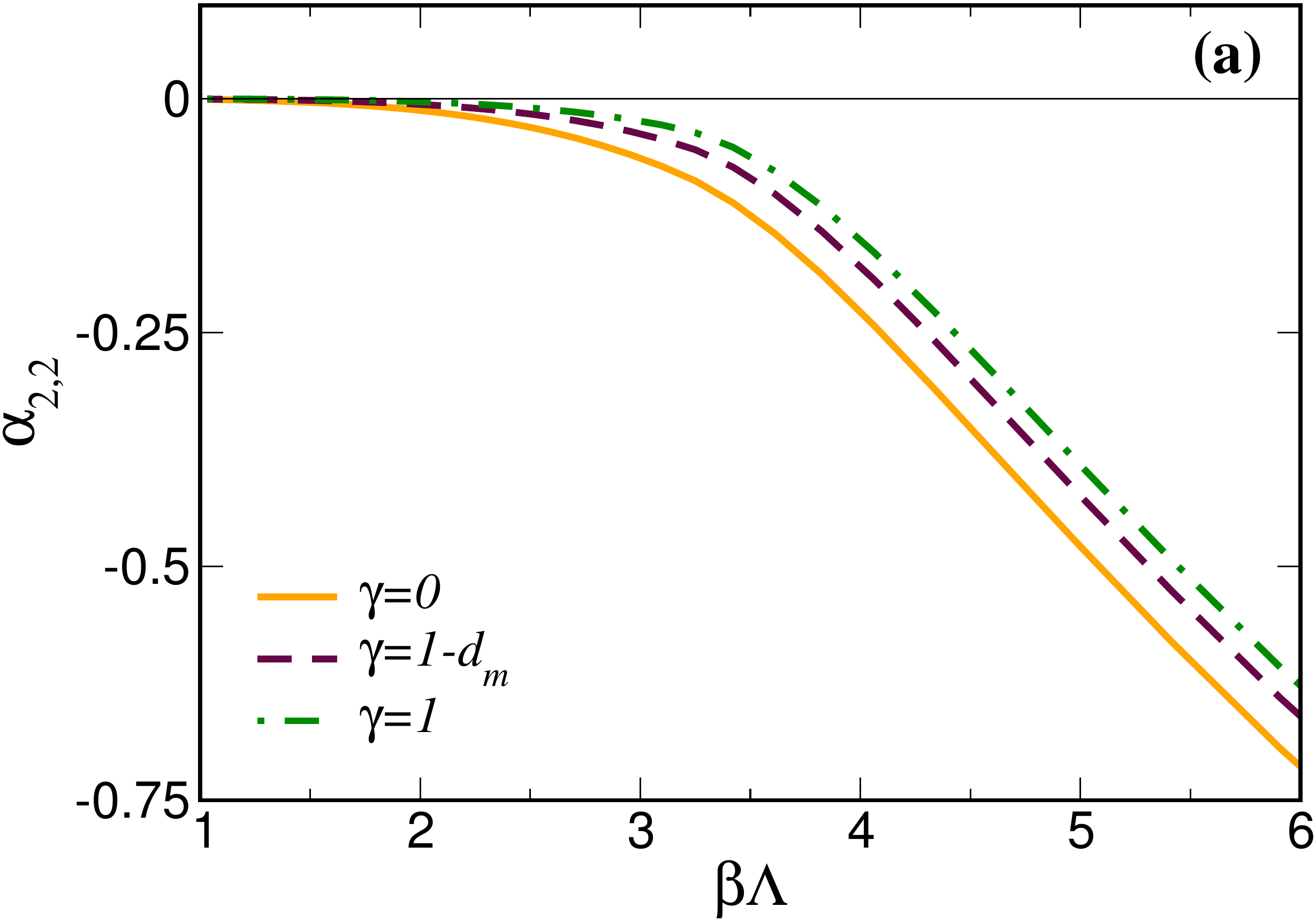}~~~~~
\includegraphics[width=7cm]{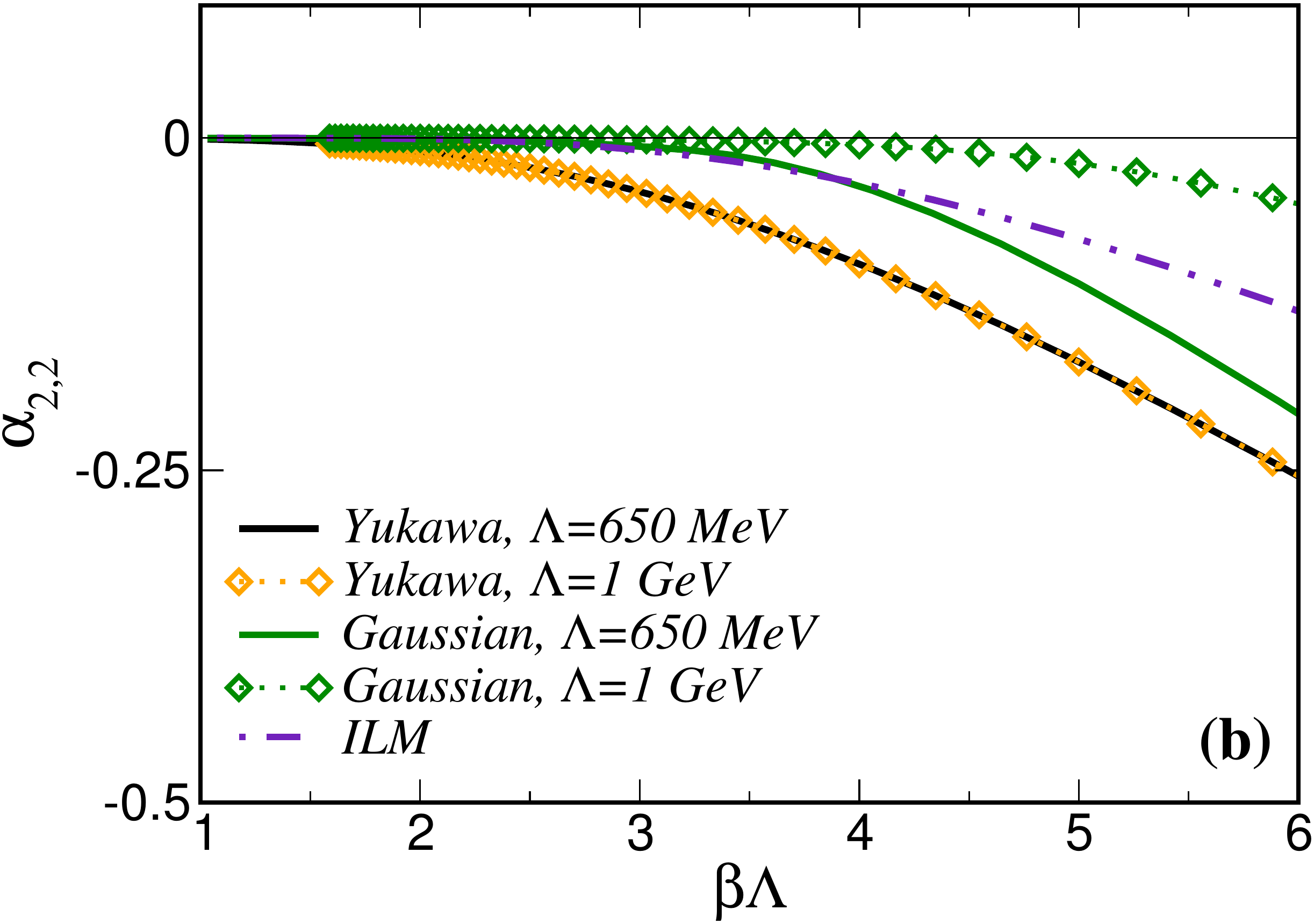}
\end{center}
\caption{\label{Fig:3}{\it (a).} Coefficient $\alpha_{2,2}$ versus $\beta\Lambda$ obtained within the 
nonlocal 4D model with the quark mass function specified by Eqs.(\er{eq:LL2}) and(\er{eq:mm2}).
{\it (b).} Coefficient $\alpha_{2,2}$ versus $\beta\Lambda$ obtained within the 
nonlocal 4D model with the quark mass function specified by Yukawa-like form factor Eq.(\er{eq:LL2y}),
nonlocal NJL form factor in Eq.(\er{eq:LL2c}) and Instanton Liquid Model form factor in Eq.(\er{eq:LL2ilm}).
For the latter we have introduced a fictitious scale $\Lambda=650$ MeV on abscissas in order to make
comparison with other models easier. }
\end{figure}
In Figure~\ref{Fig:3} we plot the coefficient $\alpha_{2,2}$ computed by Eq.(\er{eq:asd}) for the several 
form factors described in Section~\ref{Sec:2}. For all the models of $p_E-$dependent quark mass functions we find that 
$\alpha_{2,2}<0$, and because of Eq.(\er{eq:kl2}) this implies $\mu_5$ tends to increase
the critical temperature for chiral symmetry restoration within the model at hand. 
This is different from what is obtained within local models in which critical temperature has been found
to decrease with $\mu_5$, with the exception of~\cite{Ruggieri:2016cbq} where
renormalization has been used to treat the divergent vacuum term.

\subsection{The modes contributions}
It is instructive to present an analysis of the coefficient $\alpha_{2,2}$
defined in Eq.~(\ref{eq:asd}), in order to enlighten the difference between the nonlocal and local models
for what concerns $T_c(\mu_5)$ for small values of $\mu_5$. 
This analysis follows a similar one presented in \cite{Yu:2015hym} for the case
of an NJL model with a local interaction kernel and a 3-dimensional cutoff.
For simplicity, we focus on the form factor given 
by Eq.(\er{eq:LL2}) with $\gamma=0$ which allows
easier manipulations and a clearer mode separation. 
We split $\alpha_{2,2}$ as
\begin{equation}
\alpha_{2,2}={\cal I}_1 + {\cal I}_2 + {\cal J}_1 + {\cal J}_2;
\label{eq:lp2}
\end{equation}
here we have introduced several contributions depending on the momentum region of quarks 
and on temperature. These terms are defined as follows.
Firstly we add and subtract the $T=0$ contribution to  Eq.~(\ref{eq:asd}),
that according to the well known rules of finite temperature field theory in the imaginary time
formalism reads
\begin{eqnarray}
\alpha_{2,2}^0 &=& -4N_c N_f  \int\frac{d^4\bm p_E}{(2\pi)^4}{\cal C}^2(p_E)
\frac{2(3\bm p^2 - p_4^2)}{(\bm p^2 + p_4^2)^3};
\label{eq:asd_2}
\end{eqnarray}
then we define
\begin{eqnarray}
{\cal I}_1&=&-4N_c N_f  \int_{p_E^2\leq\Lambda^2}\frac{d^4\bm p_E}{(2\pi)^4}{\cal C}^2(p_E)
\frac{2(3\bm p^2 - p_4^2)}{(\bm p^2 + p_4^2)^3},\\
{\cal I}_2 &=& -4N_c N_f  \int_{p_E^2>\Lambda^2}\frac{d^4\bm p_E}{(2\pi)^4}{\cal C}^2(p_E)
\frac{2(3\bm p^2 - p_4^2)}{(\bm p^2 + p_4^2)^3},
\label{eq:M22_2}
\end{eqnarray}
which correspond to the contributions to $\alpha_{2,2}$
at zero temperature of the modes with 
$p_E^2\leq\Lambda^2$ and $p_E^2>\Lambda^2$  respectively. Moreover we define
\begin{eqnarray}
{\cal J}_1&=&-4N_c N_f T \sum_n \int
\frac{d^3\bm p}{(2\pi)^3}{\cal C}^2(\omega_n,\bm p)
\left.\frac{2(3\bm p^2 - \omega_n^2)}{(\bm p^2 + \omega_n^2)^3}
\right|_{\omega_n^2 + \bm p^2 \leq \Lambda^2}
\nonumber\\
&&~~~+4N_c N_f  \int_{p_E^2\leq\Lambda^2}\frac{d^4\bm p_E}{(2\pi)^4}{\cal C}^2(p_E)
\frac{2(3\bm p^2 - p_4^2)}{(\bm p^2 + p_4^2)^3},\\
{\cal J}_2 &=& -4N_c N_f T \sum_n \int\frac{d^3\bm p}{(2\pi)^3}{\cal C}^2(\omega_n,\bm p)
\left.
\frac{2(3\bm p^2 - \omega_n^2)}{(\bm p^2 + \omega_n^2)^3}\right|_{\omega_n^2 + \bm p^2 > \Lambda^2}
\nonumber\\
&&~~~+
4N_c N_f  \int_{p_E^2>\Lambda^2}\frac{d^4\bm p_E}{(2\pi)^4}{\cal C}^2(p_E)
\frac{2(3\bm p^2 - p_4^2)}{(\bm p^2 + p_4^2)^3},
\label{eq:M22_4}
\end{eqnarray}
which correspond to the contributions to $\alpha_{2,2}$
at finite temperature of the modes with 
$p_E^2\leq\Lambda^2$ and $p_E^2>\Lambda^2$  respectively. 

Evaluation of integrals and summation over Matsubara frequencies in the above equations
lead to the following results:
\begin{itemize}
\item modes with $p_E^2\le\Lambda^2$ at $T=0$:
\begin{equation}
{\cal I}_1=-a_2\frac{4N_c N_f}{2\pi^2}\log\frac{\Lambda}{m_0};
\label{eq:M12}
\end{equation}
\item modes with $p_E^2>\Lambda^2$ at $T=0$:
\begin{equation}
{\cal I}_2=-a_1\frac{4N_c N_f}{2\pi^2};
\label{eq:M22}
\end{equation}
\item modes with $p_E^2\le\Lambda^2$ at $T>0$:
\begin{equation}
{\cal J}_1=\frac{4N_c N_f}{2\pi^2}\left[a_2\log\frac{1}{\beta m_0} +|F(\beta\Lambda)|  \right];
\label{eq:M32}
\end{equation}
\item modes with $p_E^2>\Lambda^2$ at $T>0$:
\begin{equation}
{\cal J}_2=\frac{4N_c N_f}{2\pi^2}|G(\beta\Lambda)|;
\label{eq:M42}
\end{equation}
\end{itemize}
in order to obtain the above equations we have done some manipulation
on the definitions in Eqs.~(\ref{eq:M22_2}) and~(\ref{eq:M22_2}) which allow to extract the analytical contribution
shown in Eqs.~(\ref{eq:M12}) - (\ref{eq:M42}).
The coefficients $a_1\approx0.25$ and $a_2\approx0.938$ are the results of numerical integration.
Moreover we have introduced an infrared cutoff $m_0$ which appears in the intermediate
steps of the computation when the contributions are split; this fictitious cutoff
disappears when the sum of the contributions is done, as it is clear from Eqs.(\er{eq:M32})
and~(\eqref{eq:M12}).
In Fig.~\ref{Fig:4}a we plot the functions 
$F$, $G$ as well as their sum in order to understand the role of the several terms in Eq.(\er{eq:lp2}).
In particular the modes in Eq.(\er{eq:M22}) come from the high momentum part of the Dirac sea; 
they are not usually considered in a local model calculation because in that case their contribution is
divergent hence it is simply subtracted. We notice that this contribution to $\alpha_{2,2}$ is negative,
thus it helps to keep the critical temperature at finite $\mu_5$ higher than that at $\mu_5=0$. 

\subsection{Comparison with local NJL model}
The benefit of expansion in Eq.~(\ref{eq:lp2}) is that it allows to compare easily nonlocal with local models. 
To this end we introduce a local $\Lambda-$NJL
model in which we remove all the modes with $p_E^2 > \Lambda^2$; the coefficient 
$\alpha_{2,2}$ will be thus given by the sum of Eqs.~(\ref{eq:M12}) and~(\ref{eq:M32}) namely 
\begin{equation}
\alpha_{2,2}^{\Lambda-\mathrm{NJL}}=-\frac{4N_c N_f}{2\pi^2}
\left[
a_2\log\beta\Lambda-|F(\beta\Lambda)| 
\right].
\end{equation}
We also introduce the standard local NJL model in which we remove the ultraviolet
modes $p_E^2 > \Lambda^2$ only at $T=0$, and integrate over all momenta
at finite temperature:
\begin{equation}
\alpha_{2,2}^{\mathrm{NJL}} =-\frac{4N_c N_f}{2\pi^2}
\left[
a_2\log\beta\Lambda-|F(\infty)| 
\right],
\end{equation}
where $F(\infty)\equiv\lim_{x\rightarrow\infty}F(x)$. Both these models follow the
definitions already introduced in \cite{Yu:2015hym}.

\begin{figure}[t!]
\begin{center}
\includegraphics[width=7cm]{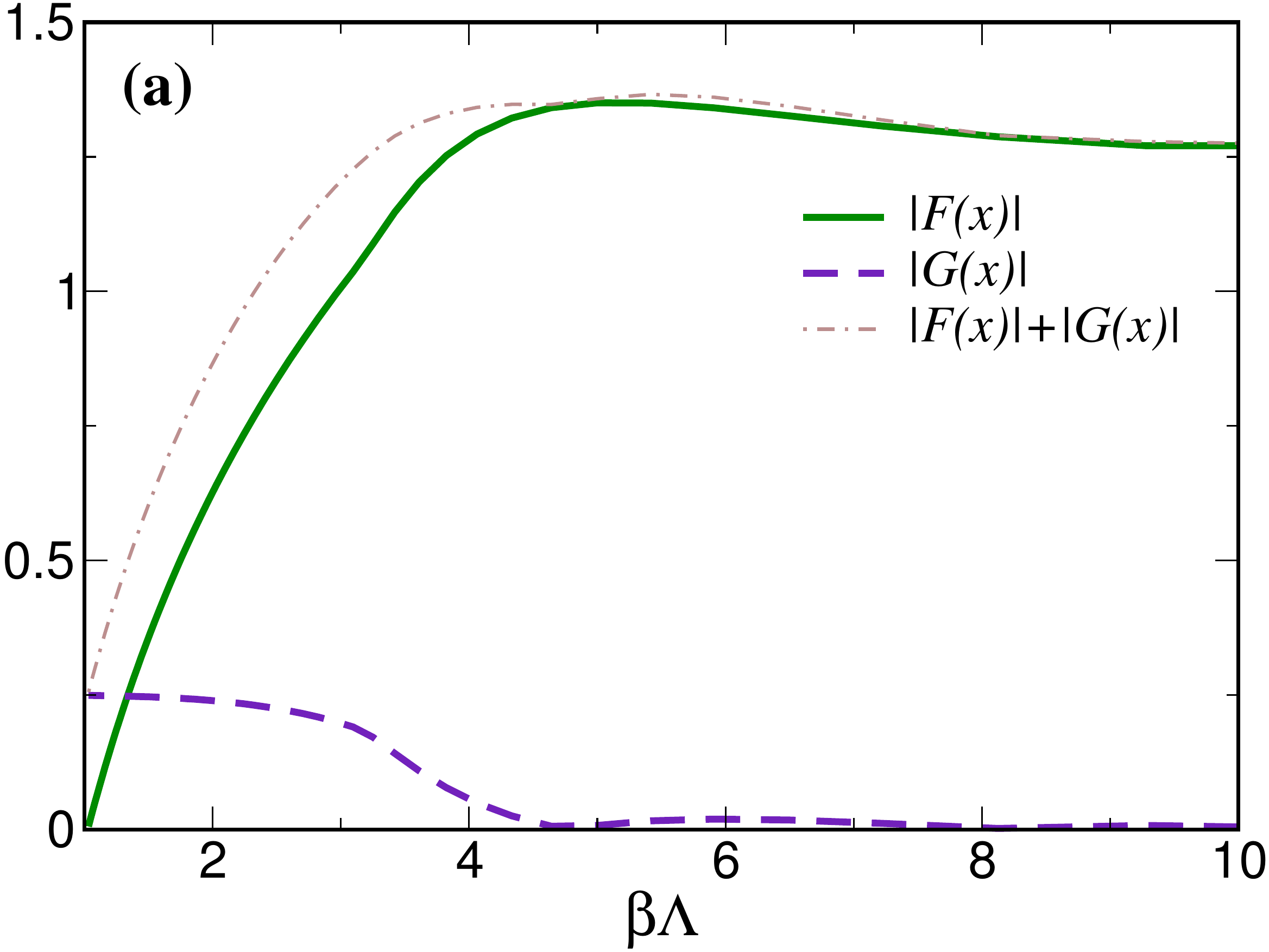}~~~~~
\includegraphics[width=7cm]{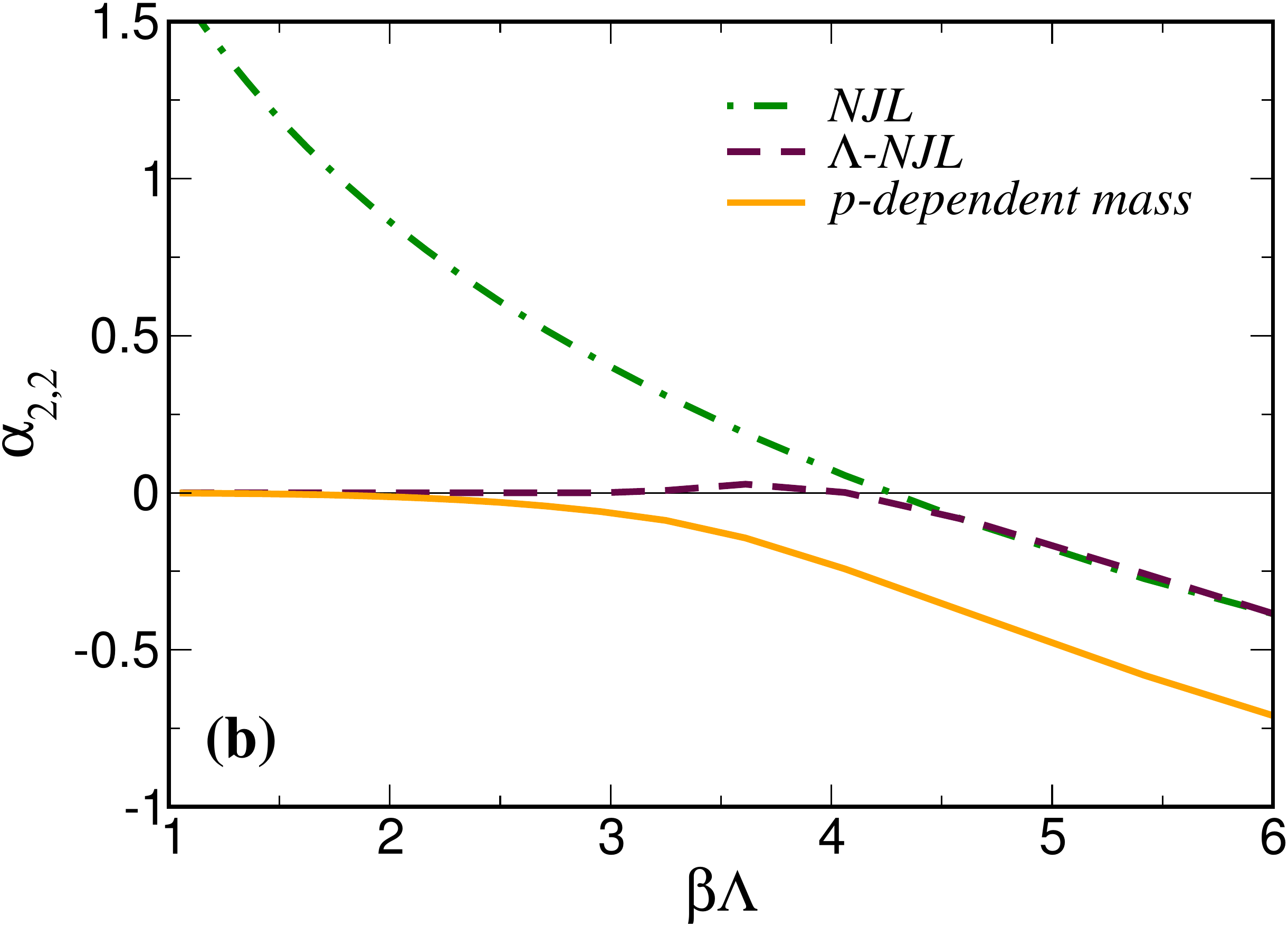}
\end{center}
\caption{\label{Fig:4}{\it (a).} Functions $F(x)$, $G(x)$ and their sum.
{\it (b).} Comparison of the $\alpha_{2,2}$ coefficients for the standard local NJL model (gren dot-dashed line),
the $\Lambda-$NJL model (maroon dashed line) and the nonlocal model with mass function
given by Eqs.(\er{eq:LL2}) and(\er{eq:mm2}) with $\gamma=0$.}
\end{figure}

In Fig.~\ref{Fig:4}b we plot the coefficient $\alpha_{2,2}$ for the 
the local NJL model (gren dot-dashed line),
the local $\Lambda-$NJL model (maroon dashed line) and the nonlocal model with mass function
given by Eqs. (\er{eq:LL2}) and (\er{eq:mm2}) with $\gamma=0$. 
For the local models there exists a window of $\beta\Lambda$ in which $\alpha_{2,2}>0$;
on the other hand for the nonlocal model considered here we find $\alpha_{2,2}<0$
for any value of $\beta\Lambda$.
The fact that $\alpha_{2,2}$  can be positive in the local models is
in part due to the absence of the vacuum term 
in Eq. (\er{eq:M22}) which would give a negative contribution to $\alpha_{2,2}$. 
Moreover, the main difference between the standard local NJL and the $\Lambda$-NJL models
is that in the latter the positive contribution ${\cal J}_2$ of the modes with $p_E^2>\Lambda^2$ at finite temperature 
is missing, while in the former the positive contribution of these modes
is added assuming a constant mass function: this explains why $\alpha_{2,2}$ 
for the $\Lambda$-NJL model is always smaller than the one of the standard local NJL.

The difference between the nonlocal model on the one hand, and the local models on the other hand,
is that for the former we find $\alpha_{2,2}$ is always negative, while for the latters there exist windows of 
$\beta\Lambda$ in which $\alpha_{2,2}$ is positive. This means that depending on the values of
$T_c$ and $\Lambda$, $T_ c$ can either increase or decrease with $\mu_5$ in local NJL models,
the result depending on model parameters.
The parameter window in which $\alpha_{2,2}$ is positive (implying $T_c$ decreasing
with $\mu_5$) is very tiny for the $\Lambda$-NJL model,
but it is quite wide for the standard local NJL model. 
Considering that for the 4-dimensional
regularization typically $\Lambda\approx 1$ GeV and $T_c$ is in the range $150-200$ MeV,
the value of $\beta\Lambda$ at $T=T_c$ turns out to be approximately in the range $5- 6.7$:
in this range we find that $\alpha_{2,2}$ is negative, which explains why in this work we find that
$T_c$ increases with $\mu_5$ also in the case of the local models.

\begin{figure}[t!]
\begin{center}
\includegraphics[width=7cm]{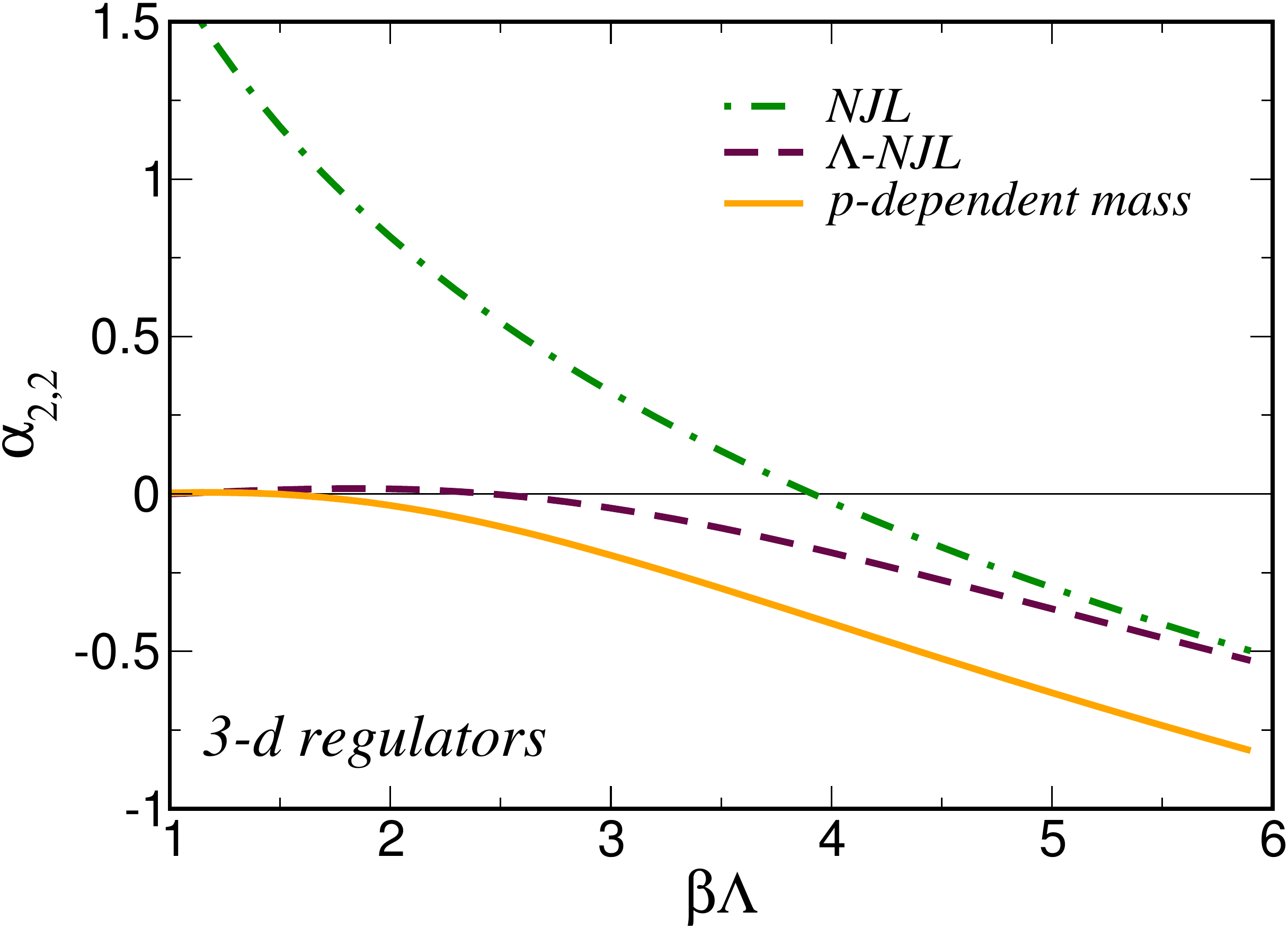}
\end{center}
\caption{\label{Fig:4bf}Comparison of $\alpha_{2,2}$ for the standard local NJL model (gren dot-dashed line),
the $\Lambda-$NJL model (maroon dashed line) and the nonlocal model with mass function
given by Eqs.(\er{eq:LL2}) and(\er{eq:mm2}) with $\gamma=0$. In all models 
we have used either a 3-dimensional regulator or a quark mass function depending on 3-momentum.}
\end{figure} 
 
In Fig.~\ref{Fig:4bf} we plot $\alpha_{2,2}$ computed for a 3-dimensional regulator;
the calculation steps are similar to those of the models with 4-dimensional regulator so we do not repeat them.
In particular the green dot-dashed line corresponds to the model used in 
\cite{Gatto:2011wc,Fukushima:2010fe,Ruggieri:2011xc}.
For 3-dimensional regularizations the value of $\Lambda$ is considerably smaller than the one 
used in the 4-dimensional case, typically
of the order of $600$ MeV \cite{Klevansky:1992qe,Hatsuda:1994pi,Buballa:2003qv}
while the range in which $T_c$ runs is the same found in the 4-dimensional case.
This implies that $\beta\Lambda$ at $T=T_c$ for 3-dimensional regularization schemes
are in the range $3-4$: for the case of the local NJL model we find this range to be in the region 
where $\alpha_{2,2}$ is positive, see green dot-dashed line in Fig~\ref{Fig:4bf}, 
meaning that $T_c$ is lowered by $\mu_5$.
Similarly for the case of the $\Lambda$-NJL model we find that 
$\alpha_{2,2}$ is negative implying that $T_c$ increases with $\mu_5$.
Comparing the results for ths standard local NJL models with 4-dimensional
and 3-dimensional regulators we thus conclude that in previous calculations 
\cite{Gatto:2011wc,Fukushima:2010fe,Ruggieri:2011xc}
the critical temperature decreases with $\mu_5$  because of an accident
driven by the model parameters, suggesting that behaviour of $T_c$ 
to be an artifact of the 3-dimensional regularization.

\section{Conclusions}
In this article we have presented a model study of the critical temperature
of chiral symmetry restoration, $T_c$, as a function of chiral chemical potential, $\mu_5$.
We have used a nonlocal NJL model with several Euclidean interaction kernels,
chosen to mimick the constituent quark mass
of QCD in the ultraviolet. 

We have studied the thermodynamic potential both within a Ginzburg-Landau expansion
in the vicinity of the second order critical line, and within calculations using the full potential.
The main interest of our study has been the computation of the critical temperature versus $\mu_5$.
The results about $T_c(\mu_5)$ are collected in Fig.~\ref{Fig:6} for the different models.
We have found that within the nonlocal models used in our study,
$T_c$ increases with $\mu_5$ regardless of the interaction kernel used. 
We remark that our interaction kernels lack of a backreaction of $\mu_5$,
hence our results should be taken with a grain of salt for $\mu_5=O(1~\mathrm{GeV})$,
while they are reliable for smaller values of $\mu_5$.
We have also found that $T_c$ increases with $\mu_5$ for a standard NJL model with a 4-dimensional
regulator, at least for small values of $\mu_5$. According to these findings, we have concluded that 
previous works \cite{Gatto:2011wc,Fukushima:2010fe,Chernodub:2011fr,Ruggieri:2011xc}
found $T_c$ a decreasing function of $\mu_5$ as a result of an accident
driven by the model parameters, suggesting that behaviour of $T_c$ 
to be an artifact of the 3-dimensional regularization of the standard local NJL model. 

We have then checked the order of the phase transition by computing the coefficient $\alpha_4$
of the GL effective potential: we have found that although $\mu_5$ makes the transition
sharper because the magnitude of $\alpha_4$ decreases with $\mu_5$ at $T_c$,
the coefficient never vanishes as it should happen at the critical endpoint. 
We have confirmed the results obtained within the GL expansion by performing a calculation
considering the full thermodynamic potential.
Our conclusion is that there is no trace of a critical endpoint in the phase diagram,
at least within the range of $\mu_5$ we have explored in this article.
According to this result we can conclude that the presence of the critical point
in the $\mu_5-T$ plane advertised before \cite{Gatto:2011wc,Fukushima:2010fe,Chernodub:2011fr,Ruggieri:2011xc}
is model dependent: in particular its existence depends on the details of the interaction used in the
model calculation. This result, as well as our conclusion about $T_c(\mu_5)$, agree with \cite{Yu:2015hym}.
 
We would like to close this article by doing few considerations about the implications
of our study. The main purpose of our investigation is purely theoretical:
the interest of a phase diagram of QCD in the $\mu_5-T$ plane was suggested
in several references \cite{Gatto:2011wc,Fukushima:2010fe,Chernodub:2011fr,Ruggieri:2011xc},
where it was found that the critical temperature for chiral symmetry restoration and for confinement-deconfinement
decrease with $\mu_5$, and a critical endpoint appears in the phase diagram.
Since both these characteristics belong also the would-be phase diagram of QCD in the
$\mu-T$ plane, and because the $\mu_5-T$ plane can be accessed by Lattice QCD calculations
while QCD at finite $\mu$ suffers the sign problem,
the idea that might derive from \cite{Gatto:2011wc,Fukushima:2010fe,Chernodub:2011fr,Ruggieri:2011xc}
is that Lattice QCD studies at finite $\mu_5$ can shed a light on QCD in the $\mu-T$ plane.
Therefore the main purpose of our model study has been to check whether the predictions 
of \cite{Gatto:2011wc,Fukushima:2010fe,Chernodub:2011fr,Ruggieri:2011xc}
are general or specific to the model used in the calculations. What we have found is that the latter scenario
is actually verified,
since classes of effective models exist in which the phase diagram looks quite different
from that advertised previously.  The scenario depicted here is in agreement with Lattice QCD 
calculations \cite{Braguta:2015owi,Braguta:2015zta},
and with results obtained by solving Schwinger-Dyson equations 
at finite $\mu_5$~\cite{Xu:2015vna,Wang:2015tia}. Therefore we can conclude by stating that 
we have now three independent calculation schemes that agree on the fact that $T_c$ increases with $\mu_5$,
and that the transition line is of the second order. As a consequence it seems unlikely that 
further investigations at finite $\mu_5$ can teach something about the QCD phase structure
in the $\mu-T$ plane.

Regarding the confinement-deconfinement in the $\mu_5-T$ plane, Lattice QCD has found no evidence
for a split of this crossover from the chiral crossover \cite{Braguta:2015owi,Braguta:2015zta}.
In order to study this problem within the models at hand we should augment the NJL model
with some physical quantity
that is sensitive to the deconfinement: the best candidate model is the NJL model augmented with 
a coupling to the Polyakov loop (PNJL)  \cite{Meisinger:1995ih,Fukushima:2003fw}. 
We expect that the picture drawn in this article does not change drastically 
by turning to the PNJL model,
in particular if a coupling between the NJL interaction and the Polyakov loop is taken into 
account \cite{Kondo:2010ts,Sakai:2010rp}.
A study of the problems studied in our article
by means of the PNJL model might be the subject of a future study.
Moreover, the absence of a critical endpoint in the $\mu_5-T$ plane might limit the 
inhomogenous condensates that have been predicted to develop in the $\mu-T$ plane
near the critical point, see for example  \cite{Nickel:2008ng,Carignano:2010ac,Abuki:2011pf,Abuki:2013pla}.
More study related to this topic might be worth of an investigation.

Finally, we would like to mention that during the very final stage of preparation of the
present manuscript, Ref.~\cite{Frasca:2016rsi} appeared in which the same problem
has been studied and an increasing $T_c$ versus $\mu_5$
has been found, in agreement with the results presented in this article. 
Moreover during the revision of the manuscript Ref. \cite{Cui:2016zqp} appeared,
in which similar conclusions to the work presented here as well as to that of \cite{Yu:2015hym}
have been drawn, regarding $T_c(\mu_5)$ and the order of the phase transition at finite $\mu_5$.

{\em Acknowledgments}. The authors would like to thank the 
CAS President's International Fellowship Initiative (Grant No. 2015PM008), 
and the NSFC projects (11135011 and 11575190). 
M.~R. also acknowledges discussions with M. Frasca.

\end{document}